\documentclass{llncs}

\usepackage{latexsym}

\def\01{\{0,1\}}
\newcommand{\ceil}[1]{\lceil{#1}\rceil} 
 
\newcommand{\eps}{\varepsilon} 
\newcommand{\ket}[1]{|#1\rangle}

\newcommand{\poly}{\mbox{\rm poly}}
\newcommand{\polylog}{\mbox{\rm polylog}}
\newcommand{\op}[1]{#1}
\newtheorem{fact}{Fact}

\newenvironment{algorithm}[1]{\medskip\noindent%
\itemsep0pt\begin{trivlist}\item[]%
{\flushleft\textbf{Algorithm: #1}}
\begin{enumerate}}%
{\end{enumerate}\end{trivlist}\medskip}

\begin{document}

\title{Quantum Search on Bounded-Error Inputs}
\author{Peter H\o yer\inst{1}\fnmsep\thanks{Supported in part
        by the Alberta Ingenuity Fund and the Pacific Institute 
        for the Mathematical Sciences.}
        \and
        Michele Mosca\inst{2}\fnmsep\thanks{Supported by St.~Jerome's University, 
        the Canada Research Chair programme, NSERC (CRO and Discovery Grant), 
        CFI, OIT, PREA, ORDCF and MITACS.}
        \and
        Ronald de~Wolf\inst{3}\fnmsep\thanks{This research
        was (partially) funded by projects QAIP
        (IST--1999--11234) and RESQ (IST--2001--37559) 
        of the IST-FET programme of the EC.}}
\institute{Dept.{} of Computer Science, Univ.{} of Calgary, Alberta, Canada.
        \email{{hoyer}\boldmath$\mathchar"40$\texttt{cpsc.ucalgary.ca}}
        \and
        Dept.{} of Combinatorics \&\ Optimization, Univ.{} of Waterloo, and
        Perimeter Institute for Theoretical Physics, Ontario, Canada.
        \email{{mmosca}\boldmath$\mathchar"40$\texttt{uwaterloo.ca}}
        \and
        CWI. Kruislaan 413, 1098 SJ, Amsterdam, the Netherlands.
        \email{{rdewolf}\boldmath$\mathchar"40$\texttt{cwi.nl}}}
\date{}
\maketitle

\begin{abstract}
Suppose we have $n$ algorithms, quantum or classical, each 
computing some bit-value {\em with bounded error probability}.
We describe a quantum algorithm that uses $O(\sqrt{n})$ repetitions 
of the base algorithms and with high probability finds the index of 
a 1-bit among these $n$ bits (if there is such an index).
This shows that it is not necessary to first significantly 
reduce the error probability in the base algorithms to $O(1/\poly(n))$
(which would require $O(\sqrt{n}\log n)$ repetitions in total).
Our technique is a recursive interleaving of amplitude amplification 
and error-reduction, and may be of more general interest.
Essentially, it shows that quantum amplitude amplification can be
made to work also with a {\em bounded-error\/} verifier.
As a corollary we obtain optimal quantum upper bounds of $O(\sqrt{N})$
queries for all constant-depth AND-OR trees on $N$ variables,
improving upon earlier upper bounds of $O(\sqrt{N}\polylog(N))$.
\end{abstract}

\section{Introduction}

One of the main successes of quantum computing 
is Grover's algorithm~\cite{grover:search,bbht:bounds}.
It can search an $n$-element space in $O(\sqrt{n})$ steps,
which is quadratically faster than any classical algorithm.
The algorithm assumes oracle access to the elements in the space,
meaning that in unit time it can decide whether the $i$th element 
is a solution to its search problem or not.  
In some more realistic settings we can efficiently make such 
an oracle ourselves.  For instance, if we want to decide satisfiability 
of an $m$-variable Boolean formula, the search space is the set
of all $n=2^m$ truth assignments, and we can efficiently decide
whether a given assignment satisfies the formula.
However, in these cases the decision is made without 
any error probability.  In this paper we study the complexity 
of quantum search if we only have {\em bounded-error\/} access
to the elements in the space.

More precisely, suppose that among $n$ Boolean values $f_1,\ldots,f_n$ we 
want to find a solution (if one exists), i.e., an index $j$ such that $f_j=1$. 
For each $i$ we have at our disposal an algorithm
$F_i$ that computes the bit $f_i$ with two-sided error: if $f_i$ is 1 then 
the algorithm outputs 1 with probability, say, at least $9/10$, 
and if $f_i=0$ then it outputs 0 with probability at least $9/10$.
Grover's algorithm is no longer applicable in this bounded-error setting, 
at least not directly, because the errors in each step will quickly 
add up to something uncontrollably large. Accordingly, we need to 
do something different to get a quantum search algorithm that works here.
We will measure the complexity of our quantum search algorithms 
by the number of times they call the underlying algorithms $F_i$.
Clearly, the $\Omega(\sqrt{n})$ lower bound for the standard error-less
search problem, due to Bennett, Bernstein, Brassard, and 
Vazirani~\cite{bbbv:str&weak},
also applies to our more general setting. 
Our aim is to give a matching upper bound.

An obvious but sub-optimal quantum search algorithm is the following.
By repeating $F_i$ $k=O(\log n)$ times and outputting the majority value
of the $k$ outcomes, we can compute $f_i$ with error probability at most $1/100n$.
If we then copy the answer to a safe place and reverse the computation
to clean up (most of) the workspace, then we get something that is 
sufficiently ``close'' to perfect oracle access to the $f_i$ bits
to just treat it as such.
Now we can apply Grover's algorithm on top of this, and because quantum
computational errors add linearly~\cite{bernstein&vazirani:qcomplexity},
the overall difference with perfect oracle access will be negligibly small.
This solves the bounded-error quantum search problem using $O(\sqrt{n}\log n)$
repetitions of the $F_i$'s, which is an $O(\log n)$-factor worse than 
the lower bound. Below we will refer to this algorithm as 
``the simple search algorithm''.

A relatively straightforward improvement over 
the simple search algorithm is the following.
Partition the search space into $n/\log^2 n$ blocks of size 
$\log^2 n$ each. Pick one such block at random. We can find a 
potential solution (an index $j$ in the chosen block such that 
$f_j=1$, if there is such a $j$) in complexity $O(\log n\log\log n)$
using the simple search algorithm, and then verify that it is indeed
1 with error probability at most $1/n$ using another $O(\log n)$ 
invocations of $F_j$.  Applying Grover search on the space 
of all $n/\log^2 n$ blocks, we obtain an algorithm with complexity 
$O(\sqrt{n/\log^2 n})\cdot O(\log n\log\log n+\log n)=O(\sqrt{n}\log\log n)$.

A further improvement comes from doing the splitting recursively:
we can use the improved upper bound to do the computation of the 
``inner'' blocks, instead of the simple search algorithm.
Using $T(n)$ to denote the complexity on search space of size $n$,
this gives us the recursion
$$
T(n)\leq d\left(T(\log^2 n)\sqrt{\frac{n}{\log^2 n}}+\log n\right)
$$
for some constant $d>0$. This recursion resolves to complexity
$O(\sqrt{n}\cdot c^{\log^*n})$ for some constant $c>0$. It is similar to
(and inspired by) the communication complexity protocol for the 
disjointness problem of H\o yer and de Wolf~\cite{hoyer&wolf:disjeq}.

Apart from being rather messy, this improved algorithm is still not optimal.
The main result of this paper is to give a relatively clean algorithm
that uses the optimal number $O(\sqrt{n})$ of repetitions to solve 
the bounded-error search problem.  
Our algorithm uses a kind of ``carrot-and-stick'' approach 
that may be of more general interest.  Roughly speaking,
it starts with a uniform superposition of all $F_i$. It then amplifies 
all branches of the computation that give answer 1. These branches 
include solutions, but they also include ``false positives'': branches 
corresponding to the $1/10$ error probability of $F_i$'s where $f_i=0$.
We then ``push these back'' by testing whether a 1-branch
is a real positive or a false one (i.e., whether $f_i=1$ or not) and 
removing most of the false ones.
Interleaving these amplify and push-back steps properly, 
we can amplify the weight of the solutions to a constant 
using $O(\sqrt{n})$ repetitions.
At this point we just do a measurement, see a potential solution
$j$, and verify it classically by running $F_j$ a few times.

As an application of our bounded-error quantum search algorithm,
in Section~\ref{sec:andor} we give optimal quantum algorithms 
for constant-depth AND-OR trees in the query complexity setting. 
For any constant $d$, we need only 
$O(\sqrt{N})$ queries for the $d$-level AND-OR tree,
improving upon the earlier $O(\sqrt{N}(\log N)^{d-1})$ 
algorithms of Buhrman, Cleve, and Widgerson~\cite{BuhrmanCleveWigderson98}.
Matching lower bounds of $\Omega(\sqrt{N})$ were already shown 
for such AND-OR trees, using Ambainis' quantum adversary method
\cite{ambainis:lowerbounds,barnum&saks:qreadonce}.
Finally, in Section~\ref{sec:general} we indicate how the ideas
presented here can be cast more generally 
in terms of amplitude amplification.

\section{Preliminaries}

Here we briefly sketch the basics and notation of quantum computation,
referring to the book by Nielsen and Chuang~\cite{nielsen&chuang:qc} for more detail.
An {\em $m$-qubit state} is a linear combination of all classical $m$-bit states
$$
\ket{\phi}=\sum_{i\in\01^m}\alpha_i\ket{i},
$$
where $\ket{i}$ denotes the basis state $i$ (a classical $m$-bit string),
the {\em amplitude\/} $\alpha_i$ is a complex number, and $\sum_i|\alpha_i|^2=1$.
We view $\ket{\phi}$ as a $2^m$-dimensional column vector.
A measurement of state $\ket{\phi}$ will give $\ket{i}$ with probability
$|\alpha_i|^2$, and the state will then collapse to the observed $\ket{i}$.
A non-measuring quantum operation corresponds 
to applying a unitary $(=$ linear and norm-preserving$)$ transformation $U$
to the vector of amplitudes.
If $\ket{\phi}$ and $\ket{\psi}$ are quantum states on 
$m$ and $m'$ qubits, respectively, then the two-register 
state $\ket{\phi}\otimes\ket{\psi}=\ket{\phi}\ket{\psi}$
corresponds to the $2^{m+m'}$-dimensional vector 
that is the tensor product of $\ket{\phi}$ and $\ket{\psi}$.

The setting of query complexity is as follows.
For input $x\in\01^n$, a {\em query} corresponds to
the unitary transformation $O$ that maps
$\ket{i,b,z}\rightarrow\ket{i,b\oplus x_i,z}$.
Here $i\in[n]$ and $b\in\01$;
the $z$-part corresponds to the workspace, which is not affected by the query.
A $T$-query quantum algorithm has the form $A=U_TOU_{T-1}\cdots OU_1OU_0$,
where the $U_k$ are unitary transformations, independent of $x$.
This $A$ depends on $x$ only via the $T$ applications of $O$.
The algorithm starts in initial all-zero state $\ket{\vec{0}}$ and its 
output (which is a random variable) 
is obtained from observing some dedicated part of the
final superposition $A\ket{\vec{0}}$.

\section{\mbox{Optimal Quantum Algorithm for Bounded-Error Search}}

In this section we describe our quantum algorithm for bounded-error search.
The following two facts generalize, respectively, the Grover search 
and the error-reduction used in the algorithms we sketched in 
the introduction.

\begin{fact}[Amplitude amplication~\cite{bhmt:countingj}]
Let $S_0$ be the unitary that puts a `-' in front of 
the all-zero state $\ket{\vec{0}}$,
and $S_1$ be the unitary that puts a `-' in front of
all basis states whose last qubit is $\ket{1}$. 
Let $A\ket{\vec{0}}=\sin(\theta)\ket{\phi_1}\ket{1}+
\cos(\theta)\ket{\phi_0}\ket{0}$ where
angle $\theta$ is such that $0 \leq \theta \leq \pi/2$ and
$\sin^2(\theta)$ equals the probability that a measurement of the
last register of state $A\ket{\vec{0}}$ yields a~'1'.
Set $G=-AS_0A^{-1}S_1$.
Then $GA\ket{\vec{0}}=
\sin(3\theta)\ket{\phi_1}\ket{1}+\cos(3\theta)\ket{\phi_0}\ket{0}$.
\end{fact}

Amplitude amplification is a process that is used in many quantum
algorithms to increase the success probability.  Amplitude
amplification effectively implements a rotation by an angle $2 \theta$
in a two-dimensional space (a space different from the Hilbert space
acted upon) spanned by $\ket{\phi_1}\ket{1}$ and
$\ket{\phi_0}\ket{0}$.  Note that we can always apply amplitude
amplification regardless of whether the angle $\theta$ is known to us
or not.

\begin{fact}[Error-reduction]
Suppose $A\ket{\vec{0}}=\sqrt{p}\ket{\phi_b}\ket{b}+
\sqrt{1-p}\ket{\phi_{1-b}}\ket{1-b}$,
where $b\in\01$ and $p\geq 9/10$. 
Then using $O(\log(1/\eps))$ applications of $A$
and majority-voting, we can build a unitary $E$ such that
$E\ket{\vec{0}}=\sqrt{q}\ket{\psi_b}\ket{b}+
\sqrt{1-q}\ket{\psi_{1-b}}\ket{1-b}$
with $q\geq 1-\eps$, and $\ket{\psi_{b/1-b}}$ possibly of larger 
dimension than $\ket{\phi_{b/1-b}}$ (because of extra workspace). 
\end{fact}

We will recursively interleave these two facts to get a quantum 
search algorithm that searches the space $f_1,\ldots,f_n\in\01$.
We assume each $f_i$ is computed by unitary $F_i$ with 
success probability at least $9/10$.
Let $\Gamma=\{j:f_j=1\}$ be the set of solutions, 
and $t=|\Gamma|$ its size (which is unknown to our algorithm).
The goal is to find an element in $\Gamma$ if $t\geq 1$,
and to output `no solutions' if $t=0$.

We will build an algorithm that has a superposition of all $j\in[n]$
in its first register, a growing second register that contains
workspace and other junk, and a 1-qubit third register indicating 
whether something is deemed a solution or not.
The algorithm will successively increase the weight of the basis
states that simultaneously have a solution in the first register 
and a 1 in the third.

Consider an algorithm $A$ that runs all $F_i$ once in superposition,
producing the state $A\ket{\vec{0}}$, which we rewrite as
$$
\frac{1}{\sqrt{n}}\sum_{i=1}^n\ket{i}\left(\sqrt{p_i}\ket{\psi_{i,1}}\ket{1}+
\sqrt{1-p_i}\ket{\psi_{i,0}}\ket{0}\right)
=
\sin(\theta)\ket{\phi_1}\ket{1}+\cos(\theta)\ket{\phi_0}\ket{0},
$$
where $p_i$ is the probability that $F_i$ outputs 1, 
the states $\ket{\psi_{i,b}}$ describe the workspace of the $F_i$,
and $\sin(\theta)^2=\sum_{i=1}^n p_i\geq 9t/10n$.

The idea is to apply a round of amplitude amplification to $A$ to 
amplify the $\ket{1}$-part from $\sin(\theta)$ to $\sin(3\theta)$.
This will amplify both the good states $\ket{j}\ket{1}$ for $j\in\Gamma$
and the ``false positives'' $\ket{j}\ket{1}$ for $j\not\in\Gamma$
by a factor of $\sin(3\theta)/\sin(\theta)\approx 3$
(here we didn't write the second register). 
We then apply an error-reduction step to reduce the amplitude of 
the false positives, setting ``most'' of its third register to 0.  
These two steps together form a new algorithm that puts 
almost 3 times as much amplitude on the solutions as $A$ does,
and that puts less amplitude on the false positives than $A$.
We then repeat the amplify-reduce steps on this new algorithm to
get an even better algorithm, and so on.

Let us be more precise. Our algorithm will consist of a number of rounds.
In round $k$ we will have a unitary $A_k$ that produces
$$
A_k\ket{\vec{0}}=
\alpha_k\ket{\Gamma_k}\ket{1}+\beta_k\ket{\overline{\Gamma}_k}\ket{1}+
\sqrt{1-\alpha_k^2-\beta_k^2}\ket{H_k}\ket{0},
$$
where $\alpha_k,\beta_k$ are non-negative reals,
$\ket{\Gamma_k}$ is a unit vector whose first register only
contains $j\in\Gamma$, $\ket{\overline{\Gamma}_k}$ is a unit vector 
whose first register only contains $j\not\in\Gamma$, 
and $\ket{H_k}$ is a unit vector. 
If we measure the first register of the above state, we will see 
a solution (i.e. some $j\in\Gamma$) with probability 
at least $\alpha_k^2$.
$A_1$ is the above algorithm $A$, which runs the $F_i$ in superposition.
Initially, $\alpha_1^2\geq 9t/10n$ 
since each solution contributes at least $9/10n$.
We want to make the good amplitude $\alpha_k$ grow by 
a factor of almost 3 in each round.

\bigskip

{\bf Amplitude amplification step.}
For each round $k$, define $\theta_k\in[0,\pi/2]$ by 
$\sin(\theta_k)^2=\alpha_k^2+\beta_k^2$.
Applying amplitude amplification ($G_k=-A_kS_0A_k^{-1}S_1$) 
gives us the state $G_kA_k \ket{\vec{0}}$, which we may write as
$$
\frac{\sin(3\theta_k)}{\sin(\theta_k)}\alpha_k\ket{\Gamma_k}\ket{1}+
\frac{\sin(3\theta_k)}{\sin(\theta_k)}\beta_k\ket{\overline{\Gamma}_k}\ket{1}+
\sqrt{1-\left(\frac{\sin(3\theta_k)}{\sin(\theta_k)}\right)^2(\alpha_k^2+\beta_k^2)}\ket{H_k}\ket{0}.
$$
We applied $A_k$ twice and $A_k^{-1}$ once,
so the complexity goes up by a factor of 3.

\bigskip

{\bf Error-reduction step.}
Conditional on the qubit in the third register being 1, 
the error-reduction step $E_k$ now does majority voting on $O(k)$ 
runs of the $F_j$ (for all $j$ in superposition) to decide
with error at most $1/2^{k+5}$ whether $f_j=1$.
It adds one 0-qubit as the new third register and maps 
(ignoring its workspace, which is added to the second register) 
$$
\begin{array}{lcl}
E_k\ket{j}\ket{1}\ket{0}   & = & a_{jk}\ket{j}\ket{1}\ket{1}+\sqrt{1-a_{jk}^2}\ket{j}\ket{1}\ket{0}\\
E_k\ket{j}\ket{0}\ket{0}   & = & \ket{j}\ket{0}\ket{0}
\end{array}
$$
where $a_{jk}^2\geq 1-1/2^{k+5}$ if $f_j=1$ and 
$a_{jk}^2\leq 1/2^{k+5}$ if $f_j=0$. This way, 
$E_k$ removes most of the false positives.

\bigskip
\bigskip

\noindent
Putting $A_{k+1}=E_kG_kA_k$ and defining $\alpha_{k+1}$, $\beta_{k+1}$,
$\ket{\Gamma_{k+1}}$,  $\ket{\overline{\Gamma}_{k+1}}$, 
and $\ket{H_{k+1}}$ appropriately, we now have
$$
A_{k+1}\ket{\vec{0}}=\alpha_{k+1}\ket{\Gamma_{k+1}}\ket{1}+
\beta_{k+1}\ket{\overline{\Gamma}_{k+1}}\ket{1}+\sqrt{1-\alpha_{k+1}^2-\beta_{k+1}^2}\ket{H_{k+1}}\ket{0}.
$$ 
Here the second register has grown by the workspace used 
in the error-reduction step $E_k$, as well as by the qubit 
that previously was the third register.
The good amplitude has grown in the process:
$$
\alpha_{k+1}\geq\alpha_k\frac{\sin(3\theta_k)}{\sin(\theta_k)}
\sqrt{1-1/2^{k+5}}.
$$
Since $x-x^3/6\leq \sin(x)\leq x$, we have 
$$
\frac{\sin(3\theta_k)}{\sin(\theta_k)}\geq  3-9\theta_k^2/2.
$$
Accordingly, as long as $\theta_k$ is small, 
$\alpha_k$ will grow by a factor of almost 3 in each round. 
On the other hand, the weight of the false positives goes down rapidly:
$$
\beta_{k+1}\leq\beta_k\frac{\sin(3\theta_k)}{\sin(\theta_k)}\frac{1}{\sqrt{2^{k+5}}}.
$$
We now analyze the number $m$ 
of rounds that we need to make the good amplitude large.
In general, we have $\sin(\theta_k)^2=\alpha_k^2+\beta_k^2$,
hence $\theta_k^2\leq 2(\alpha_k^2+\beta_k^2)$ for the domain we are interested in. 
Here $\alpha_k^2\leq 9^{k-1}\alpha_1^2$ and 
$\beta_k^2\leq \frac{1}{10}(9/2^6)^{k-1}$.
Note 
\begin{eqnarray*}
\sum_{k=1}^{m-1}\theta_k^2 & \leq & 2\sum_{k=1}^{m-1} \alpha_k^2+\beta_k^2\\
 & \leq & 2\sum_{k=1}^{m-1} 9^{k-1}\alpha_1^2 + 
          2\sum_{k=1}^{m-1} \frac{1}{10}(9/2^6)^{k-1} \\
 & \leq    & 2\cdot 9^{m-1}\alpha_1^2 + 1/4.
\end{eqnarray*}
Therefore, $m$ rounds of the above process 
amplifies the good amplitude $\alpha_k$ to
\begin{eqnarray*}
\alpha_m & \geq & \alpha_1\prod_{k=1}^{m-1} 
\frac{\sin(3\theta_k)}{\sin(\theta_k)}\sqrt{1-1/2^{k+5}}\\
 & \geq & \alpha_1\prod_{k=1}^{m-1} \left(3-9\theta_k^2/2\right)
\left(1-1/2^{k+5}\right)\\
 & = & \alpha_1 3^{m-1}\prod_{k=1}^{m-1}\left(1-3\theta_k^2/2\right)
\left(1-1/2^{k+5}\right)\\
 & \geq & \alpha_1 3^{m-1}\left(1-\frac{3}{2}\sum_{k=1}^{m-1}\theta_k^2-\sum_{k=1}^{m-1}\frac{1}{2^{k+5}}\right)\\
 & \geq & \alpha_1 3^{m-1}\left(1-\frac{3}{2}(2\cdot 9^{m-1}\alpha_1^2+1/4)-1/16\right)\\
 & \geq & \alpha_1 3^{m-1}\left(1/2-3\cdot 9^{m-1}\alpha_1^2\right).
\end{eqnarray*}
In particular, whenever the (unknown) number $t$ of solutions lies 
in the interval $[n/9^{m+1},n/9^m]$, equivalently $9^m\in[n/9t,n/t]$, 
then we have 
$$
\frac{1}{3^m\sqrt{10}}\leq \sqrt{\frac{9t}{10n}}
\leq\alpha_1\leq\sqrt{\frac{t}{n}}\leq\frac{1}{3^m}.
$$
This implies 
$$
\alpha_m\geq 0.04,
$$ 
so the probability of seeing a solution 
after $m$ rounds is at least $0.0016$. By repeating this classically
a constant number of times, say 1000 times, we can bring the success 
probability close to 1 (note to avoid confusion: these 1000 
repetitions are not part of the definition of $A_m$ itself).

The complexity $C_k$ of the operation $A_k$, in terms of number 
of repetitions of the $F_i$ algorithms, is given by the recursion
$$
C_1=1\mbox{ and }C_{k+1}=3C_k + O(k),
$$
where the $3C_k$ is the cost of amplitude amplification and 
$O(k)$ is the cost of error-reduction.
This implies 
$C_m=O(\sum_{k=1}^{m-1}k\cdot 3^{m-k-1})=O(3^m).$

We now give the full algorithm when the number of solutions is unknown:

\begin{algorithm}{Quantum search on bounded-error inputs}
\item for $m=0$ to $\ceil{\log_9(n)}-1$ do:
\begin{enumerate}
\item run $A_m$ 1000 times
\item verify the 1000 measurement results,
each by $O(\log n)$ runs of the corresponding $F_j$
\item if a solution has been found, then output a solution and stop
\end{enumerate}
\item Output `no solutions'
\end{algorithm}
This finds a solution with high probability if one exists.
The complexity is
$$
\sum_{m=0}^{\ceil{\log_9(n)}-1} 1000\cdot O(3^m)+1000\cdot O(\log n)=O(3^{\log_9(n)})=O(\sqrt{n}).
$$
If we know that there is at least one solution but we don't know how
many there are, then, using a modification of our algorithm 
as in~\cite{bbht:bounds}, we can find a solution 
using an expected number of repetitions in
$O(\sqrt{N/t})$, where $t$ is the (unknown) number of solutions.  This
is quadratically faster than classically, and optimal for any quantum
algorithm.

\section{Optimal Upper Bounds for AND-OR Trees}
\label{sec:andor}

A $d$-level {\em AND-OR tree\/} on $N$ Boolean variables is a 
Boolean function that is described by a depth-$d-1$ tree with interleaved 
ORs and ANDs on the nodes and the $N$ input variables as leaves. 
More precisely, a 0-level AND-OR tree is just an input variable,
and if $f_1,\ldots,f_n$ all are $d$-level AND-OR trees on $m$ variables,
each with an AND (resp.~OR) as root, then OR$(f_1,\ldots,f_n)$ (resp.~AND) 
is a $(d+1)$-level AND-OR tree on $N=nm$ variables.
AND-OR trees can be converted easily into OR-AND trees and vice versa
using De Morgan's laws, if we allow negations to be added to the tree.

Consider the two-level tree on $N=n^2$ variables 
with an OR as root, ANDs as its children, 
and fanout $n$ in both levels.
Each AND-subtree can be quantum computed by Grover's algorithm 
with one-sided error using $O(\sqrt{n})$ queries
(we let Grover search for a `0', and output 1 if we don't find any),
and the value of the OR-AND tree is just the OR of those $n$ values.
Accordingly, the construction of the previous section
gives an $O(\sqrt{n}\cdot\sqrt{n})=O(\sqrt{N})$ 
algorithm with {\em two}-sided error.
This is optimal up to a constant factor~\cite{ambainis:lowerbounds}.

More generally, for $d$-level AND-OR trees we can apply the above 
algorithm recursively to obtain an algorithm with $O(c^{d-1}\sqrt{N})$ queries.
Here $c$ is the constant hidden in the $O(\cdot)$ of 
the result of the previous section.
For each fixed $d$, this complexity is $O(\sqrt{N})$,
which is optimal up to a constant factor~\cite{barnum&saks:qreadonce}. 
It improves upon the $O(\sqrt{N}(\log N)^{d-1})$ algorithm given 
in~\cite{BuhrmanCleveWigderson98}. 

Our query complexity upper bound also implies that the minimal
degree among $N$-variate polynomials approximating AND-OR 
is $O(\sqrt{N})$~\cite{bbcmw:polynomials}.
Whether this upper bound on the degree is optimal remains open.
The best known lower bound for the 2-level case
is $\Omega(N^{1/4}\sqrt{\log N})$~\cite{shi:restrictions}.

\section{Amplitude Amplification with Imperfect Verifier}\label{sec:general}

In this section we view our construction in a more general light.

Suppose we are given some classical randomized algorithm~$\op{A}$
that succeeds in solving some problem with probability~$p$. 
In addition, we are given a Boolean function $\chi$
that takes as input an output from algorithm $\op{A}$, and outputs
whether it is a solution or not.  Then, we may find a solution to our
problem by repetition.  We first apply algorithm $\op{A}$, obtaining
some candidate solution, which we then give as input to the
verifier~$\chi$.  If $\chi$ outputs that the candidate indeed is a
solution, we output it and stop, and otherwise we repeat the process
by reapplying $\op{A}$.  The probability that this process terminates
by outputting a solution within the first $\Theta(\frac{1}{p})$
iterations of the loop, is lower bounded by a constant.

A~quantum analogue of boosting the probability of success is to boost
the \emph{amplitude} of being in a certain subspace of a Hilbert
space.  Thus far, amplitude amplification~\cite{bh:aa} has assumed
that we are given a perfect verifier~$\chi$: whenever a candidate
solution is found, we can determine with certainty whether it is a
solution or not.  Formally, we model this by letting $\chi$ be
computed by a deterministic classical
subroutine or an exact quantum subroutine.

The main result of this paper may be viewed as an adaptation
of amplitude amplification to the situation where the verifier
is {\em not\/} perfect, but sometimes makes mistakes.  
Instead of a
deterministic subroutine for computing~$\chi$, we are given a
bounded-error randomized subroutine, and instead of an
exact quantum subroutine, we are given a bounded-error quantum
subroutine.  Previously, the only known technique for handling such
cases has been by straightforward simulation of a perfect verifier:
construct a subroutine for computing $\chi$ with error $\frac{1}{2^k}$
by repeating a given bounded-error subroutine of order $\Theta(k)$
times and then use majority voting.  Using such direct simulations, we
may construct good but sub-optimal quantum algorithms, like the 
$O(\sqrt{n}\log n)$ query algorithm for quantum search of the introduction.
Here, we have introduced a modification of the amplitude amplification
process that allows us to efficiently deal with imperfect verifiers.
Essentially, our result says that imperfect verifiers are as good as
perfect verifiers (up to a constant multiplicative factor in the complexity).

\subsection*{Acknowledgments}
We thank Richard Cleve for useful discussions, 
as well as for hosting MM and RdW at the University of Calgary, 
where most of this work was done.


\end{document}